# Topology and Chirality

C. Felser
*Max Planck Institute for Chemical Physics of Solids,*
*01187 Dresden, Germany*
*E-mail: Claudia.Felser@cpfs.mpg.de*
*www.cpfs.mpg.de*

J. Gooth
*Max Planck Institute for Chemical Physics of Solids,*
*01187 Dresden, Germany*
*E-mail: Johannes.Gooth@cpf.mpg.de*
*www.cpfs.mpg.de*

Topology, a well-established concept in mathematics, has nowadays become essential to describe condensed matter. At its core are chiral electron states on the bulk, surfaces and edges of the condensed matter systems, in which spin and momentum of the electrons are locked parallel or anti-parallel to each other. Magnetic and non-magnetic Weyl semimetals, for example, exhibit chiral bulk states that have enabled the realization of predictions from high energy and astrophysics involving the chiral quantum number, such as the chiral anomaly, the mixed axial-gravitational anomaly and axions. The potential for connecting chirality as a quantum number to other chiral phenomena across different areas of science, including the asymmetry of matter and antimatter and the homochirality of life, brings topological materials to the fore.

## 1. Introduction

Electronic properties of solids play a central role in our everyday life. One recent research area, topology, became a major direction in condensed matter physics, solid state chemistry and materials science. It has led to a fundamental new understanding of solids mainly due to relativistic effects in compounds made of heavier elements [1]. Before topology in 2005 turned to the center stage, it was assumed that the electronic properties of superconductors, metals, insulators and semiconductors are completely described by the energy-momentum relations of the electrons in them, the so-called “band structures”, which in turn are defined by the symmetries of the underlying crystal lattice of the host solid. However, this description turned out to be incomplete with the prediction of the quantum spin Hall effect in 2005 [2,3], which may be viewed as the quantum Hall effect without an external magnetic field, but with strong spin-orbit coupling instead. In the Quantum Hall regime, that is when electrons are restricted to two dimensions and are exposed to a strong magnetic field, condensed matter systems can exhibit completely different electrical properties without any additional symmetries being broken [4]. Due to the quantization of electronic states in the magnetic field, such a system has an energy gap at the Fermi energy. Despite this gap, this state is not a conventional insulator, but has metallic chiral edges and a quantized Hall conductivity. Observing such a system for the first time, the German physicist Klaus von Klitzing received the Nobel Prize in 1985. This was the beginning of the topological classification of solids. However, it took more than 25 years and new predictions [2,3] before the concept of topology became a main stream direction in solid state research and was associated to an intrinsic material property: the Berry curvature — a quantum mechanical property that is related to the phase of electrons wavefunction and is particularly relevant in condensed matter systems exposed to strong magnetic fields or strong spin-orbit coupling. Meanwhile, all non-magnetic inorganic compounds are characterized by the topology of their band structures based on group and graph theory based on a single particle picture [5,6] as well as some of the most import antiferromagnetic materials classes [7]. Surprisingly, it was found that more than 20% of all materials that we know today have electronic properties that are governed by the topology of their electron wave functions, i.e. Berry curvature effects.

One of the most intriguing consequences of the topological description of materials are chiral states on the bulk, surfaces and edge of topological condensed matter systems. This intimate connection goes even far beyond the characterization of electron states and comprises today (quasi-)particle states in condensed matter systems, such as chiral electron states, chiral photons, chiral magnons, chiral plasmons, chiral spins, to name a few of them. In the context of electrons in solids, chirality $\chi$ is defined in reciprocal (momentum) space as the handedness of the spin of the electrons relative to their direction of motion. Right-handed electrons have a chirality of $\chi = -1$ and left-handed electrons have a chirality of $\chi = +1$, see Fig.1 a.

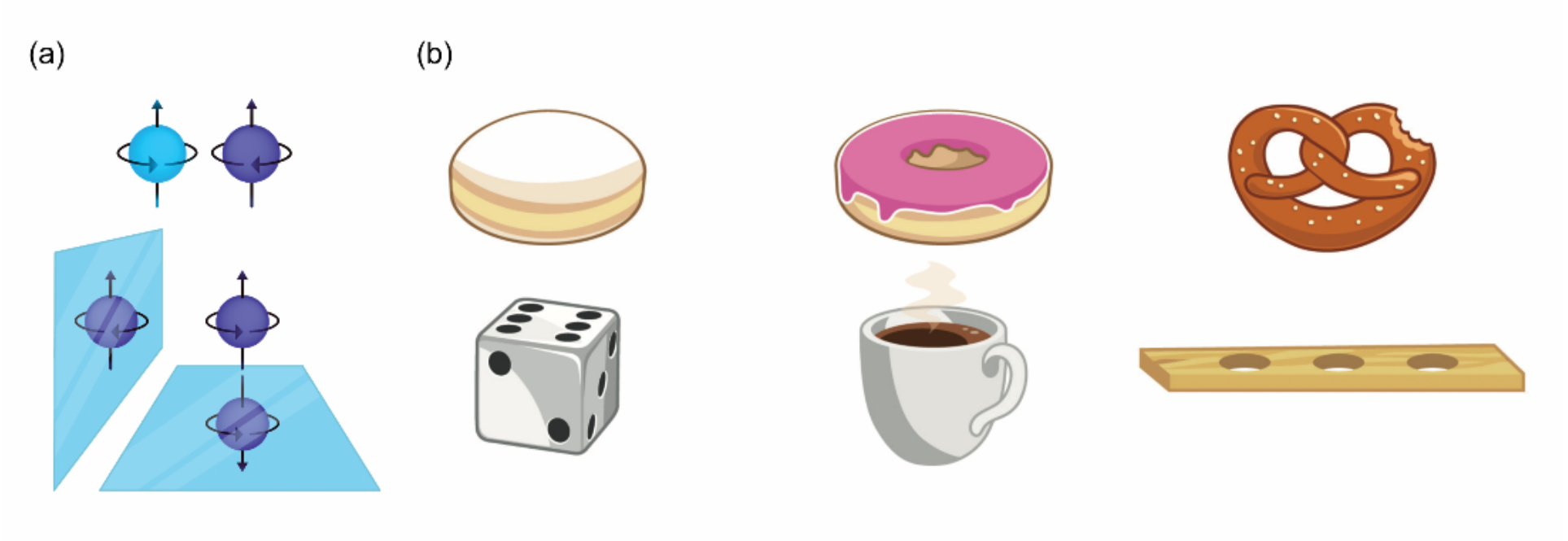

Fig. 1. (a) Electrons are chiral, if spin and momentum are locked. (b) Topology is a simple concept, dealing with the surfaces of objects. The topology of a mathematical structure is identical if it is preserved under continuous deformation. A pancake has the same topology as a cube, a donut as a coffee cup and a pretzel as a board with three holes.

The concept of chirality is an overarching theme in physics, chemistry and biology, permeating much of modern science. It links the properties of the universe and its constituent elementary particles, through organic stereochemistry, to the structure and behavior of the molecules of life, with much else besides (chemical crystallography, chiroptical spectroscopy, nonlinear optics, nanoscience, materials, electrical engineering, planar plasmonic metamaterials, spintronics, molecular motors, pharmaceuticals, astrobiology, origin of life, etc.). However, the connections between the different concepts of chirality (an overview over a few definitions of chirality in various contexts is given in box 1) all over the sciences remains elusive. This article gives the authors' personal perspective on how chirality in topological materials is potentially related to chirality as a general concept.

**Keypoints and definitions in the context of topology and chirality:**

| | |
|---|---|
| 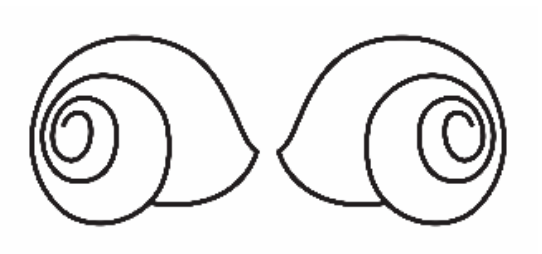 | **Chiral object:** has a geometric property of being non-superposable on its mirror image; such an object has no symmetry elements of the second kind (a mirror plane, a center of inversion, a rotation-reflection axis). [8]. |
| 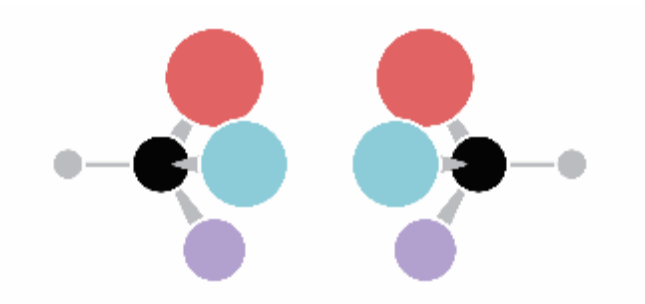 | **Chiral molecules**: are chiral, if the point group contains only symmetry operations, of the first kind (rotation and translation). Both enantiomers consist of identical chemical compositions but are distin-guished in their light–matter interaction and catalytic reactivity. Chiral molecules can crystallize in chiral and non-chiral crystal structures [9]. |
| 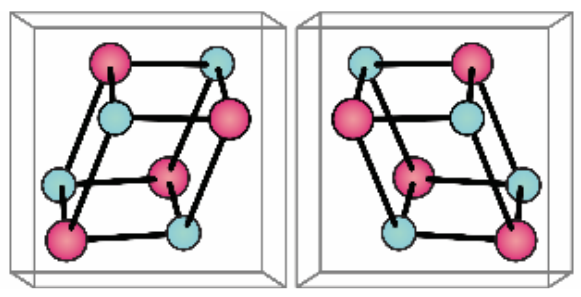 | **Chiral space groups:** contain symmetry operations of the first kind. There are 11 pairs of enantiomorphic space groups (e.g. $P6_1$ and $P6_5$) which are chiral. 43 achiral space groups can host a chiral crystal structure. A crystal structures in space group $P2_13$ for example, both enantiomeric crystals crystallize in the same space group. |
| 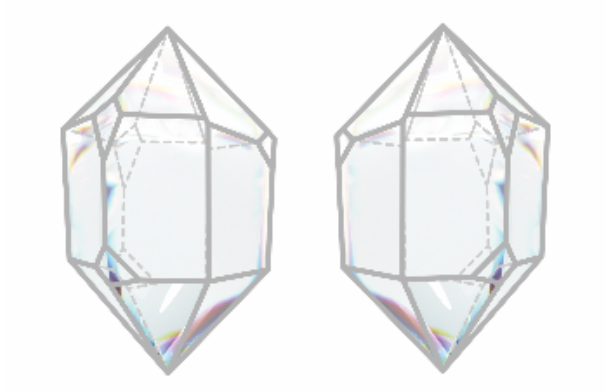 | **Chiral crystals:** is an inorganic or organic material crystallizing in one of the 65 Söhncke symmetry space group. If the space group contains only proper operations, the crystal is chiral and obey a well-defined handedness. Of all inorganic compounds in the inorganic database ICSD approximately 20% are chiral [16]. Recently, it was observed that spin-polarized currents in chiral crystals can propagate over tens of micrometers [10]. |
| 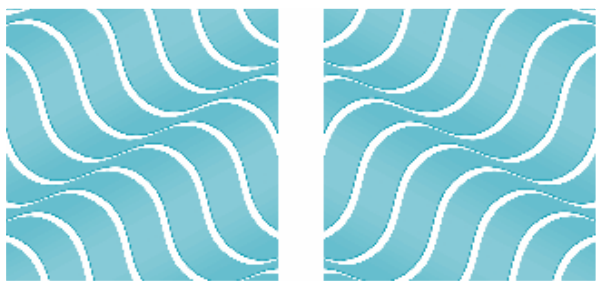 | **Chiral crystal surface**: a crystal with a surface is periodic in only in two dimensions, one consequence is that the inversion symmetry is lost. Surfaces of chiral crystals are intrinsically chiral, achiral crystal have chiral surfaces, if its surface normal does not lie in any of the mirror planes of the bulk crystal lattice; and are typically high-Miller-index surfaces [11]. |
| 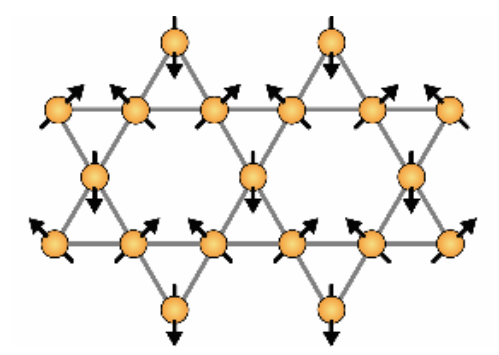 | **Chiral magnetic structure:** Magnetism breaks time reversal symmetry. In helical magnets inversion symmetry is broken, these magnets are abundant, in metals alloys, semiconductors and multiferroics. $Mn_3Z$ (Ge, Sn, Ir, Pt) were identified to be non-collinear antiferromagnets with a chiral spin arrangement and a large anomalous Hall effect (AHE) arising from the topologically non-trivial spin texture [12-14]. |
| 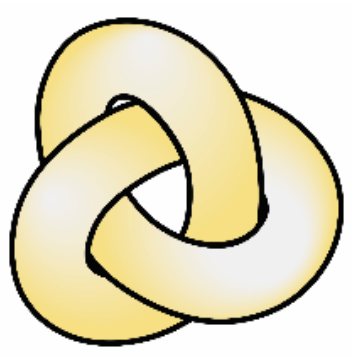 | **Topology**: is a mathematical description of the properties that are preserved through deformations. The concept in condensed matter includes symmetry, and relativistic effects such as spin orbit coupling (SOC) and band inversion [1]. All chiral crystals with large SOC are Kramers – Weyl Fermions, with monopols and antimonopols. B20 compounds have a maximal Chern equal 4 [15-17]. |
| 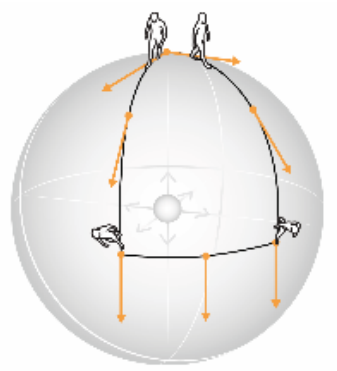 | **Berry phase:** is a geometric phase in a quantum mechanical system, when the system does not return to its initial. It gives rise to observable effects such as the anomalous Hall effect and the orbital magnetization [18]. A new concept proposes a connection of the sign of the Berry curvature with the orbital angular momentum around Weyl points in semimetals and in chiral new Fermions [19]. |
| 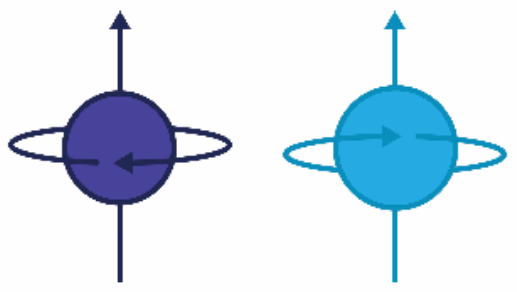 | **Spin momentum locking:** can be observed in topological insulator surface states, the spin is locked at a right angle to their momentum [20]. In chiral systems (chiral molecules) the spin is coupled to the electron linear momentum, the origin of the CISS effect [21]. |
| 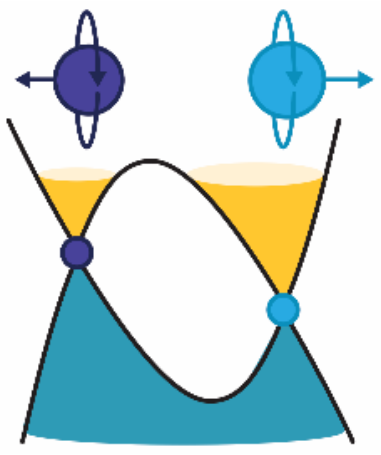 | **Chiral anomaly:** is subject of research in high-energy, condensed matter, and nonequilibrium physics via parity-breaking of chiral currents along a magnetic field [22]. It can be broken in a quantum world, in a quark-gluon plasma created in heavy-ion collisions, Floquet systems, and non-Hermitian systems. In solids the chiral anomaly can be observed in Weyl systems, when an electric field is applied parallel to a magnetic field, and charges are pumped between two Weyl points of opposite chirality [23-25]. |
| 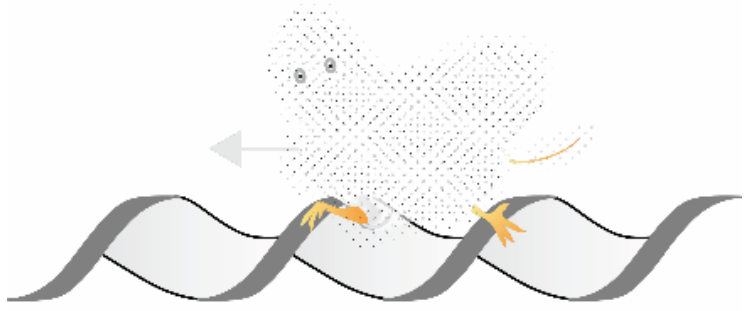 | **Chiral quasiparticle:** The quasiparticle concept was developed by Landau to describe Fermions (electrons or holes) interacting with other particles. Bosons (phonons or plasmons) are named collective excitations. Quasiparticles or collective excitations, which behave more like non-interacting particle are easier to describe. Graphene can be described by chiral quasiparticles [26]. |

## 2. Topology

Topology is the branch of mathematics that deals with the deformation of objects. Objects that can be continuously transformed into one another belong to a certain category — they are said to have a certain topological order, which is characterized by a so-called topological invariant. In contrast to this, objects that can only be reshaped by cutting or breaking into one another have different topological orders. A classic example of topological classification is by the number of holes in an object. A cup and a donut have the same topological order, but a Brezel has a different one, see Fig.1 b. Such objects can also be quantum mechanical. Electron states can be classified topological on the basis of a property that is the sum of the so-called "Berry curvature" along a closed path in momentum space. Admittedly, it is a very abstract quantity that sounds quite constructed at first. The “Berry curvature” is a quantum mechanical property that is closely linked to the phase of the electron wave function, which an electron picks up after having traveled a closed path when it has returned to the origin. This phase is called the “Berry Phase” [18]. It may sound a little intuitive that the property of an object changes when it returns to its starting point after having traveled a closed path, but systems with such properties also exist in the classical world. A well-known example is the Foucault pendulum, whose direction of oscillation deviates after one day from the value of the previous day. The same applies to the “Berry Phase”, which defines the topology of the electronic states in a solid. Solid-state crystals have the same topological order if the sum of the “Berry curvature” of their electrons along a closed path in momentum space is the same and solid-state crystals, for which this sum is different, have a different topological order. It turns out that different topological orders in solid-state crystals, despite their abstract nature, are associated with different optical and electrical properties that can actually be observed in experiments. Historically, it was believed that only a change in the system’s symmetry (that is the lattice symmetry or time-reversal symmetry) can lead to different optical and electrical properties. The significance of these different properties for the interpretation of the quantum Hall effect described above for example is that electronic states of different topological orders can merge into one another by “breaking” the band structure by means of the magnetic field without breaking an additional symmetry in the solid-state system. In other words, the quantized Hall conductivity in a two-dimensional electron system that is exposed to a strong magnetic field represents the topological invariant of the system and thus marks the topological order. A two-dimensional quantum Hall system is therefore also called a topological insulator. A characteristic property of topological insulators is the presence of metallic, i.e. dissipationless, chiral states at the sample boundary, in which the spin and momentum of the electrons is locked. Such states always occur at the spatial interface between regions that are in different topological orders. The easiest way to understand this is to note that different topological orders in insulators are associated with the parity of band gap, that defines which particular band is the conduction and which is the valance band. Imagine now a smooth boundary between two systems of opposite band parity (in one of the systems band A is the conduction band and band B is the valance band, and in the second system band B is the conduction band and band A is the valance band). At this boundary, the band structure slowly interpolates as a function of position between the two systems. Somewhere along the way, the energy gap has to disappear; otherwise both sides would be in the same class. Dissipationless chiral states are therefore bound to the interface. The surface of a quantum Hall system or a solid-state crystal can be viewed as the interface to the vacuum, which, like a conventional insulator, belongs to the trivial topological class. This guarantees the existence of gapless states on the surface (or edge) of a topological insulator. One of the most important discoveries in recent years is that topological order also occurs in some three-dimensional (3D) materials. In these materials the role of the magnetic field of a 2D-qantum Hall system is taken over by the mechanism of spin-orbit coupling, like in the 2D quantum spin Hall systems. These materials have been called 3D topological insulators, because they are insulators inside them, but because of the topological order they have exotic chiral 2D metallic surface states. Topological states of matter are so far identified via a simple single electron picture, which explains also the fast success of the field in condensed matter physics.

Recently the 3D-Quantum Hall effect was observed in a single crystal of $ZrTe_5$ [27]. Bertrand Halperin had proposed long ago in 1987 that it should be possible to realize the QHE in a three-dimensional semimetal or doped semiconductor with a particular instability in the Fermi surface [28]. The considerable challenge was to realize a single crystal, which fulfills all necessary conditions for such a 3D QHE: namely, high mobility, an adjusted charge carrier concentration and impurity level with the Fermi level tuned to be in an energy gap to the applied magnetic field. In two-dimensional electron gas (2DEG) systems, the charge carrier concentration, the mobilities and the Fermi energy can be varied by gating. Tantalizing signatures of the 3D QHE were seen in step-like anomalies in the Hall conductivity in the extreme quantum limit in Bismuth and graphite, and in some semimetallic compounds with the extreme quantum limit reached at low magnetic fields. However, no convincing plateaus as in the classical 2DEG nor finite conductivity perpendicular to the QHE were realized. Last year, however, strong evidence for a 3D QHE was found in single crystals of $ZrTe_5$ in small magnetic fields of just 2 Tesla. The electron density modulation in the direction of the magnetic field of the $ZrTe_5$ single crystals was accounted for by a charge density wave instability. The results were soon reproduced by other groups and, moreover, evidence for a 3D version of the fractional QHE was seen in single crystals of $HfTe_5$, see Fig. 2 [29].

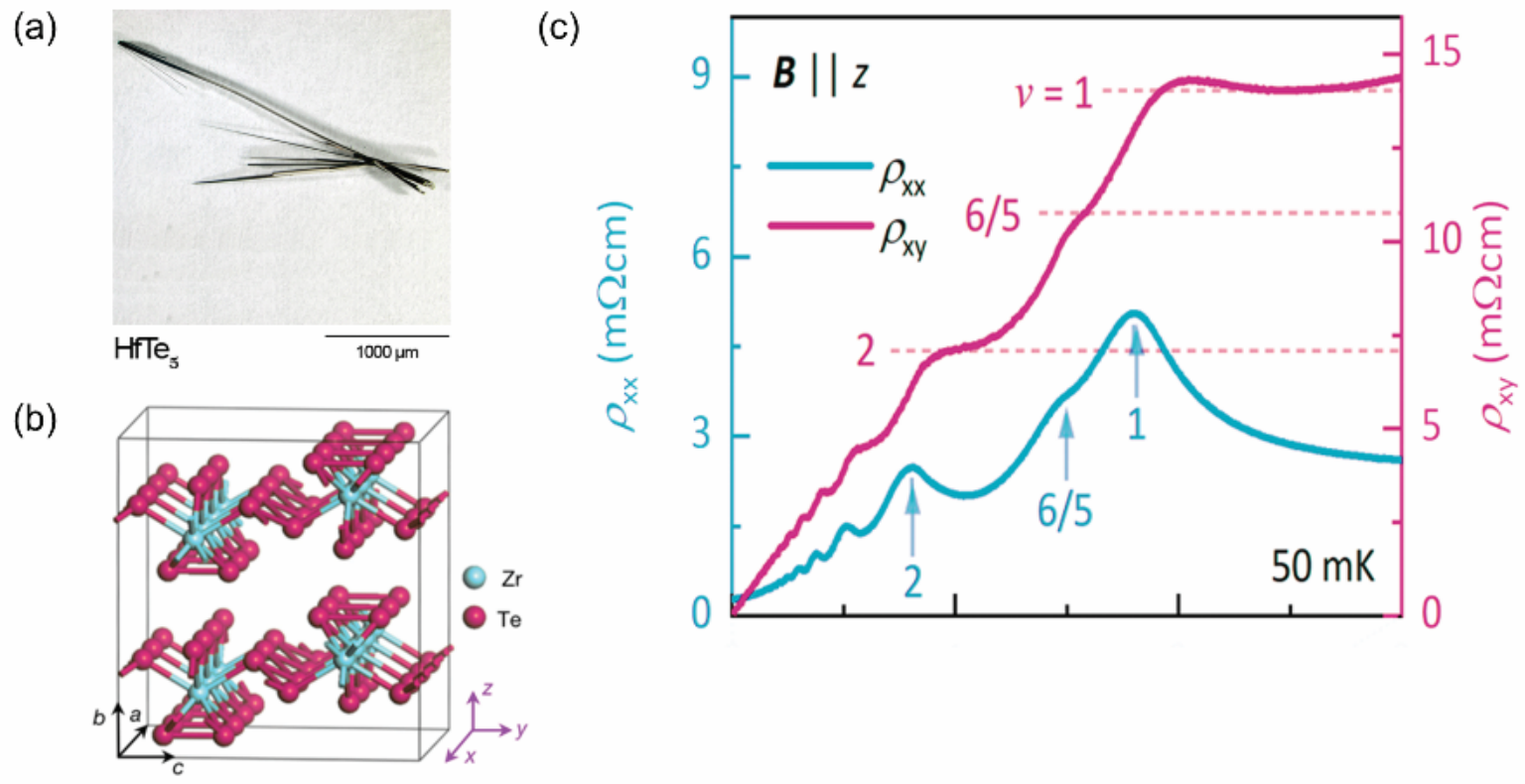

Fig. 2. Figure according to [29], (a) $HfTe_5$ crystals (b) $ZrTe_5$ and $HfTe_5$ crystal structure and (c) Longitudinal electrical resistance $\rho_{xx}$ (left axis) and Hall resistance $\rho_{xy}$ (right axis) as a function of B at 2 K with B applied along the *b*-axis of single crystal $HfTe_5$.

As already noted, although related, trivial and topological materials cannot be distinguished directly from the band structure itself. This requires the information from which atomic orbitals the individual bands come from. From a chemical point, in trivial insulating and semiconducting materials, the *s*-electrons usually form the conduction band, while the *p*-electrons form the valence band (Fig. 3 a). If the materials consist of heavy elements (atomic number $Z > 32$), the bands of the outer *s*-electrons are lower in energy compared to light elements, so that they can appear below or just at the Fermi energy. This can result in the minimum of the 5 *s*- or 6 *s*-bands of the electrons being below the maximum of the *p*-bands of the electrons, and *s*- and *p*- bands to intersect at points of intersection or nodal lines in three dimensions. One then speaks of band inversion. However, due to strong spin-orbit coupling, such intersection points and node lines can be forbidden and split (Fig. 3 b). An energy gap is created again, with parts of the *s*-electrons now forming the conduction band, while parts of the *p*-electrons also form the valence band. With regard to the band gaps of trivial isolators, one speaks in this case of a negative or inverted band gap. Materials that have such inverted band gaps are topological insulators (TI). The inversion of the 6*s*-states of Hg / Bi and 5*p*-states of Te in HgTe / $Bi_2Te_3$ are prominent examples of such inverted bands with a negative band gap. The electrons are close to the nucleus and do not like to be ionized. Examples of topological elements are bismuth from group 5, as well as phosphorus. Phosphorus can have a formal valence of +5 in phosphates, for example, while bismuth only occurs as $Bi^{3+}$, the outer *s*-electrons are not available as valence electrons. Many of the heavy element compounds exhibit unusual electronic and optical properties due to the topology of their electronic structure, e.g. one finds extremely high values or anomalies in certain experiments that can only be explained with the help of the topology. In fact, one can find metallic surfaces with three-dimensional topological insulators (Fig. 3 b, right side). In this case it is the chiral two-dimensional surfaces of the solid-state crystal that arise for the same reasons as for quantum Hall systems.

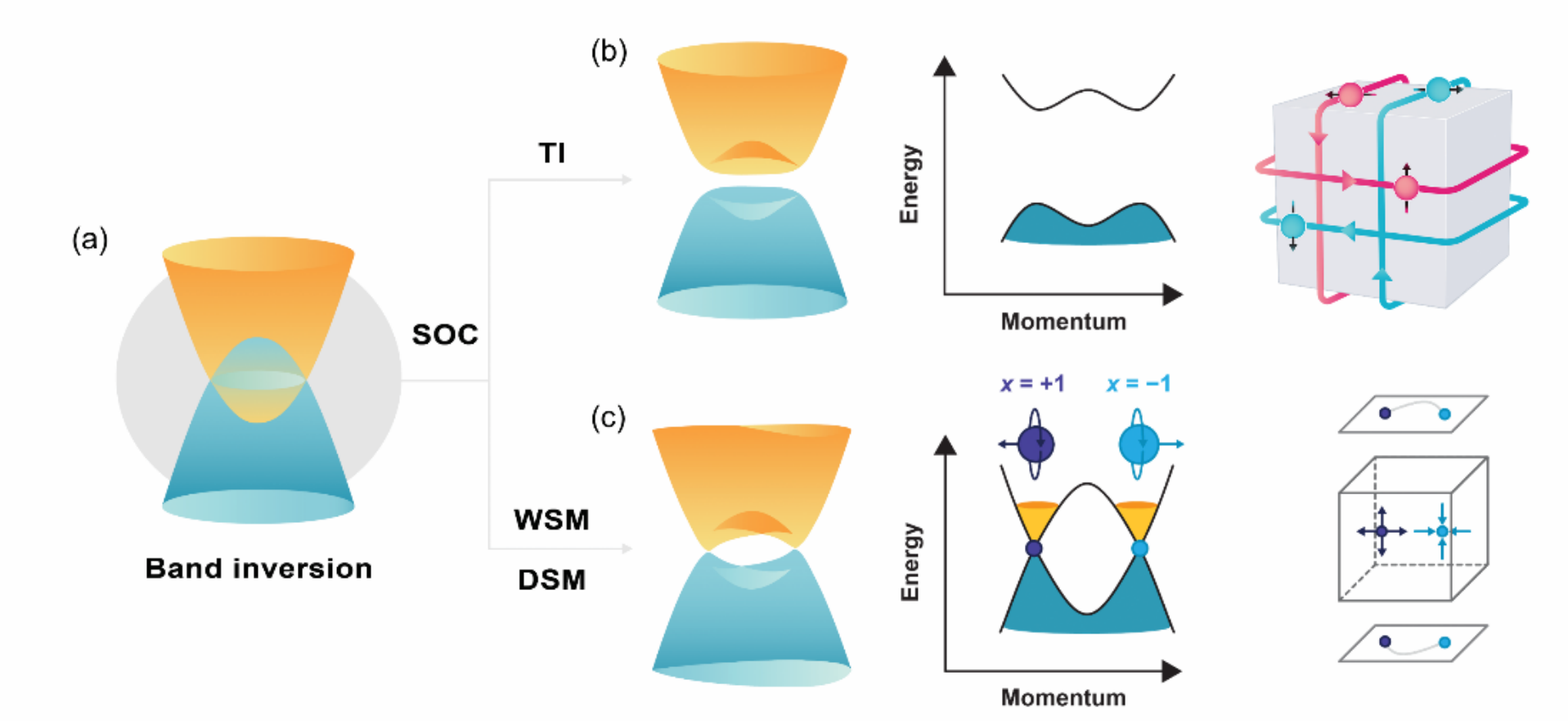

Fig. 3. Figure according to [31], displaying the topological classification of solids. The conduction band often *s*-electrons are colored yellow and those of the valence band often *p*-electrons are blue. (a) The band inversion, overlap between conduction and valence electrons, leads to a nodal line. Strong spin-orbit coupling can lead to the splitting of the bands (b). The resulting band gap is inverted and represents a topological isolator (TI), sketch of the surface states, connecting valence and conduction band (dotted lines) and spin–momentum locked surface electrons for the two spin directions on the surface of a crystal (c) sketch of a Dirac semimetal (DSM) and a Weyl semimetal (WSM). The band structure of a Weyl semimetal consists of pairs of linear crossing points with chirality $\chi = +1$ and $\chi = -1$. (b) These crossing points behave like magnetic monopole and anti-monopole in momentum space, from which special topological ones Form surface conditions, so-called Fermi arcs, which connect the Weyl crossing points.

## 3. Weyl semimetals

So far, we have limited our discussion to isolators and semiconductors. However, the electronic structure of a solid can also have crossings of inverted bands, which are stable if they are not prohibited for reasons of symmetry. Such material systems are topological semimetals, in which exotic properties are derived from the crossing points of the electron bands (Fig. 3 c). Ordinary metals and semimetals mostly always have a curved — mostly parabolic — dispersion, which describes non-relativistic electrons with finite rest mass. In contrast to this, topological semimetals in the vicinity of their crossing points often show a linear relationship between energy and momentum. Such a linear dispersion describes relativistic electrons without rest mass, which move at about a thousandth of the speed of light — just like the elementary particles in high-energy physics.

Topological semimetals with 4-dimensional crystal symmetry are called Dirac semimetals (DSM, Fig. 3 c). Graphene, a layer of carbon atoms in graphite, is the most prominent 2D example of such a material, but they also exist as three-dimensional materials. The electrons in such Dirac semimetals the energy-momentum relation is linear and the electrons behave like relativistic Dirac Fermions without rest-mass. This is in contrast to normal metals, which exhibit parabolic energy-momentum relations that describe massive electron (quasi-)particles. About 5 years ago, a new class of materials with new topological properties was identified — the so-called Weyl semimetals (WSM, Fig. 3 c), named after the physicist Hermann Weyl. At first glance, the electronic structure of a Weyl semimetal looks like that of a Dirac semimetal, with intersecting linear bands in the electronic structure of the solid. However, Dirac and Weyl semimetals differ in their symmetry and therefore also in their properties. A Dirac semimetal is centro-symmetrical, while a Weyl semimetal has a lower two-dimensional symmetry. It has no inversion center in the crystal structure — which breaks the inversion crystal symmetry, or a ferromagnetic order — which breaks the time-reversal symmetry. In contrast to Dirac semimetals, Weyl semimetals exhibits pairs of non-degenerate band crossings. The electrons at these crossing points behave like a special solution of the Dirac equation for massless particles — the so-called Weyl Fermions [22,30,31]. Weyl Fermions are characterized by the unique property that their intrinsic angular momentum or spin is inextricably linked with their linear momentum. What is special about Weyl materials is that in each pair of their crossing points, there is one crossing point with left and one crossing point with right-handed electrons. As such, the crossing points in a Weyl semimetal behave like magnetic monopole and anti-monopole in momentum space (Berry phase). The chiral volume properties result in special topological surface states, the Fermi arcs that connect the Weyl points (Fig. 3 c).

The relativistic equations that describe these linear dispersions have one mathematically inevitable feature: negative energy solutions (Fig. 4). The interpretation of these filled vacuum states of negative energy in high-energy physics has always been a bit controversial, as it contradicts our intuition of an “empty” vacuum. However, their physical reality is beyond question, as one of their direct consequences — the existence of antimatter — is confirmed by experiments. On the other hand, filled states of negative energy are a very natural concept from the point of view of solid-state physics: filled valence bands in the electronic structure of topological semimetals. The upper cone is the conduction band and represents the analogue to matter, the lower one is the valence band and represents the antimatter analogue.

As an example of a Weyl semimetal from our work, we discuss briefly NbP, a Weyl semimetal, which is characterized by breaking the inversion symmetry. We found all the typical characteristics of a Weyl semimetal in single crystals of NbP. For example, some electrons behave as if they were almost massless, which leads to extremely high mobility and a large magneto resistance. The key signature of a Weyl semimetal is evidence of a Fermi arc. As with a camera, the electronic structure and thus the sought-after Fermi arc can be photographed using angle-resolved photo emission (ARPES) [32,33].

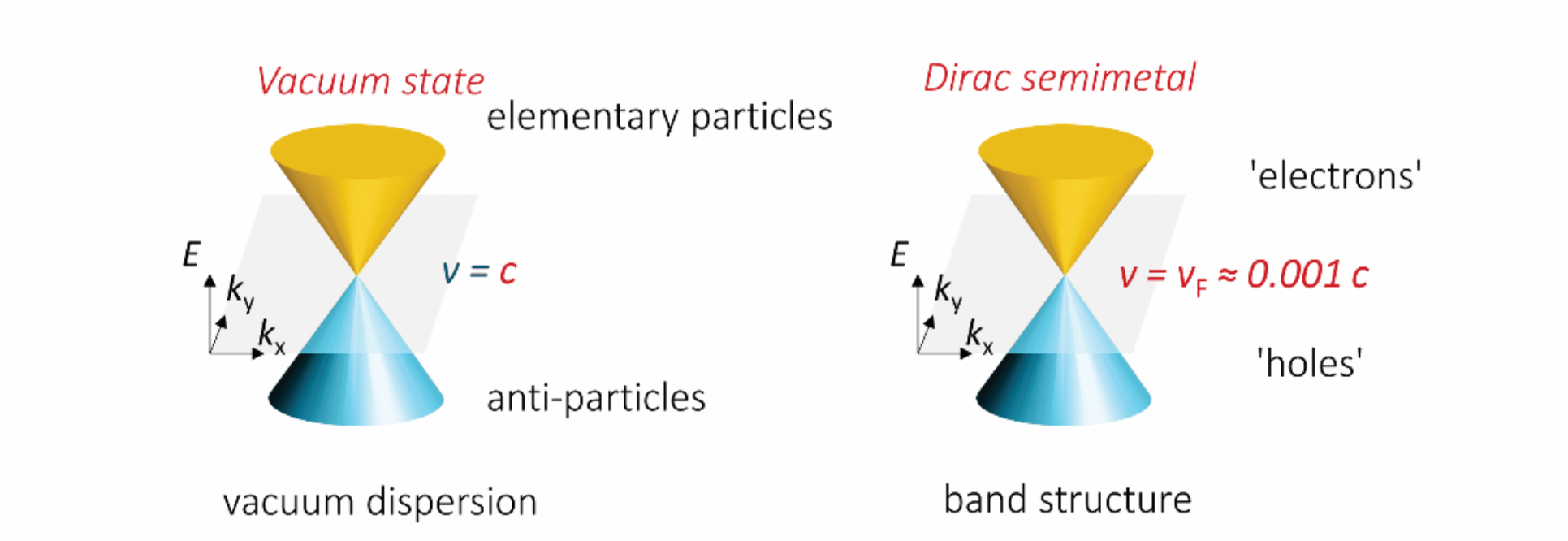

Fig. 4. In a quantum system the charge parity can be broken (charge parity violation). Comparison of the vacuum state with an imbalance between particles and anti-particles and a semimetal with an imbalance of electrons and holes, as a consequence of the chiral anomaly.

## 4. Chiral and axial-gravitational anomaly in Weyl semimetals

Classically, the chirality is a strictly conserved physical quantity — such as angular momentum, energy or electrical charge. Chirality must be preserved in the sum*, i.e.* there cannot simply be more particles of one chirality than of the other. In the context of an accelerator experiment in the 1970s, however, it was discovered that the conservation law of chirality is broken in parallel electric and magnetic fields at the quantum level. The observed decay of a neutral pion into two photons should actually be prohibited by the conservation law of chirality, but it turned out that this is not the case. Even if, in this case, it is not a matter of massless Fermions, but of bosons, the chirality is defined — however, as for massless Fermions, it can no longer be equated with the handedness of the particles. In 1969, this "chiral anomaly" was explained by the theorists Stephen Adler [34] as well as John Stewart Bell and Roman Jackiw [35] independently of one another by going from the classical to a quantum field theoretical description. Therefore, the chiral anomaly is now also known as the Adler-Bell-Jackiw anomaly. In the standard model of particle physics, certain proportions of the particle-antiparticle asymmetry are believed to be attributed to such quantum anomalies, the violation of classical conservation laws due to quantum fluctuations. The chiral anomaly, in particular, is said to play an important role in the standard model of particle physics, but Weyl Fermions have not yet been detected as elementary particles. In addition, an underlying curved spacetime is predicted to make a significant contribution to the chiral imbalance. This effect is known as the mixed axial-gravitational anomaly. In extreme gravitational fields, which correspond to a strongly curved spacetime, the axial-gravitational anomaly might thus also contribute to the particle-antiparticle asymmetry.

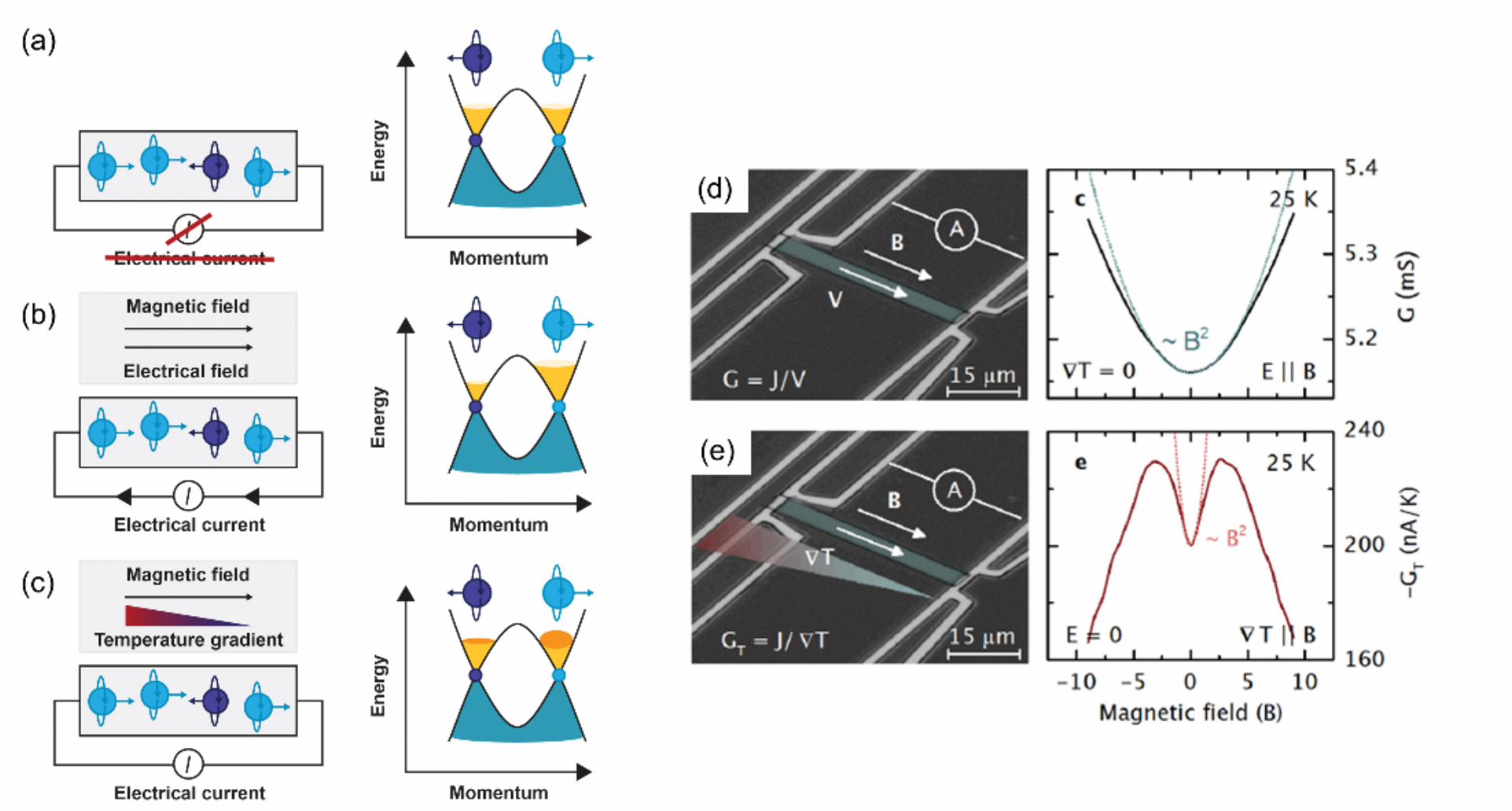

Fig. 5. Chiral and axial-gravitational anomaly in Weyl semimetals. (a) Circuit sketch (right) and band structure diagram of a Weyl semimetal (right) without a magnetic field. (b) Circuit sketch (right) and band structure diagram of a Weyl semimetal (right) in a parallel electric and magnetic field. (c) Circuit sketch (right) and band structure diagram of a Weyl semimetal (right) in parallel temperature gradient and magnetic field Figure (a-c) according to [36]. Figure (d) and (e) according to [38]. Experiments on the chiral and axial-gravitational anomaly in the Weyl semimetal NbP. Colored optical recordings of the measuring arrangement of the electric current I as a response to the electric (V) and magnetic field (B). (b) Magnetic field dependence of the electric current for different angles (color legend) between V and B. (d) Colored optical recordings of the measuring arrangement of the electric current I as a response to the temperature gradient (ΔT) and magnetic field (B). The red and green ends of the color gradient indicate the hot and cold side of the device. (e) Magnetic field dependence of the electric current for different angles (color legend) between ΔT and B.

In Weyl semimetals, the conservation law of chirality leads to a suppression of the electric current. This can be understood using a simple Weyl band structure: Figure 5 a shows a pair of Weyl crossing points on the right side, with the left crossing point representing electrons moving to the left and the right crossing point moving to the right — represents moving electrons. The spin of the electrons is the same for both crossing points. Every electron at the left crossing point has a chirality of $\chi$ = +1 and every electron at the right crossing point has a chirality of $\chi$ = – 1. In thermal equilibrium, *i.e.* when no electrical energy flows through the Weyl semimetal, the Fermi energy (the filling level of the cones) is the same in both Weyl cones. There are just as many electrons moving to the right as there are electrons moving to the left — so there is no net charge transport and, therefore, just as many right-handed as left-handed Weyl Fermions. The chirality of the entire system is zero. If you apply an electric field to a normal metal, e.g. from left to right, an electric current flows from left to right, because there are more electrons moving to the right. But not so in a Weyl semimetal. A current flow from left to right would mean a higher filling level in the right Weyl cone than in the left and thus to the generation of more right-handed Weyl Fermions in the semimetal than left-handed. However, according to the conservation law, this is prohibited. Only the application of an additional magnetic field parallel to the electric field, which is so high that it quantizes the electron states (Figure 5b), leads, analogously to high-energy physics, to the breaking of the law of conservation of chirality and enables a constant redistribution of electrons between the left and right Weyl cones . This induced breaking of the chiral symmetry is a macroscopic manifestation of the chiral anomaly in the relativistic field theory and leads in Weyl semimetals to a positive longitudinal magnetic field-dependent electric current [36]. However, care has to be taken in experiments, because a positive magnetoconductance can also be cause by extrinsic effects, such as by a magnetic field-induced inhomogeneous current distribution inside the sample [37]. Samples with lower mobilities, well defined shape and additional cross-check experiments, like probing the mixed axial-gravitational anomaly in thermal and thermoelectric experiments performed under open electrical circuit conditions (no net charge current flow) are required to exclude such extrinsic effects. One might think that since the gravitational fields on our earth are relatively weak and experiments on Weyl semimetals always take place in relatively flat spacetime, that the mixed axial-gravitational anomaly does not play a role in solid-state physics. In fact, however, the mixed axial-gravitational anomaly leads to measurable effects in magnetic field-dependent thermal and thermoelectric experiments through a “back door”. You have to know that temperature gradients in relativistic systems act like a gravitational field. This goes back to the American theorist Luttinger, who introduced statistical quantities such as entropy and temperature into quantum field theory in 1964. At a similar time, Tolman and Ehrenfest, who argued that temperature is not constant is space at thermal equilibrium, but varies with curved space time. Simply put, one can say that mass and energy are the same in relativistic systems. As is well known, masses are moved in gravitational fields and energy as heat in temperature gradients. Consequently, temperature gradients in relativistic systems represent analogues for gravitational fields. If a temperature gradient is applied to a normal metal, e.g. from left (warm) to right (cold), an electric current flows from left to right because the hot electrons diffuse from left to right. But not so in a Weyl semimetal. Corresponding to an electric field, a current flow from left to right would mean a higher filling level in the right Weyl cone than in the left and thus for the generation of more right-handed Weyl Fermions in the semimetal than left-handed. However, according to the conservation law, this is prohibited. Only the application of an additional magnetic field parallel to the temperature gradient, which is so high that it quantizes the electron states (Figure 5c), leads, analogous to high-energy physics, to the breaking of the chirality conservation law and enables a constant redistribution of electrons between the left and right Weyl cones. This induced breaking of the chiral symmetry is a macroscopic form of the mixed axial-gravitational anomaly in the relativistic field theory and leads to a positive longitudinal magnetic field-dependent thermoelectric current in Weyl semimetals. We were able to use these relationships to demonstrate the chiral (Figure 5d) and axial-gravitational anomaly (Figure 5e) in the Weyl semimetal NbP by measuring the electrical and thermoelectric currents in the magnetic field in 2017 [38]. The relative orientation of the electric field or temperature gradient to the magnetic field is decisive in these experiments. Both the chiral and axial-gravitational anomaly are experimentally reflected in an electrical current that increases with the magnetic field; this is only the case if the magnetic field and the electric field current are parallel to one another, otherwise the electric and thermoelectric current will decrease as the magnetic field increases. A critical component in these experiments is the ability to synthesize high quality samples. Therefore, groups capable of synthesizing high-quality single crystal solids are leaders in this field.

In case of chiral particles, not only the chiral symmetry itself, but also the energy-momentum tensors of the chiral quasiparticles are separately conserved. The energy-momentum tensor encodes the density and flux of energy and momentum, i.e. measures the contributions of quasiparticle currents and heat currents to the total energy currents in the system. However, in strong gravitational fields, in addition to the conservation law of chirality also this conservation law of the momentum tensors should be violated by quantum fluctuations. The separate conservation of the energy-momentum tensors of the chiral particles represents another consequence of the gravitational anomaly, but is elusive today. It is the emergent separate conservation at the Weyl cones at Weyl semimetals that makes us confident that it will be possible to probe the gravitational anomaly in the thermal transport of Weyl semimetals in the future exposed. In such systems, the gravitational anomaly is expected to cause a positive contribution to the longitudinal magneto-thermal conductivity in such systems. In thermal equilibrium, both left and right-movers have the same temperature, and are, hence, described by a similar energy-momentum tensor. If a temperature gradient is applied to a normal metal, e.g. from left (warm) to right (cold), a heat current flows from left to right because the hot electrons diffuse from left to right. But not so in a Weyl semimetal. A heat current flow from left to right would mean a higher temperature in the right Weyl cone than in the left and thus hotter right-handed Weyl Fermions in the semimetal than left-handed ones. However, according to the conservation law, this is prohibited. Only the application of an additional magnetic field parallel to the temperature gradient, which is so high that it quantizes the electron states, leads, analogous to the predictions high-energy physics, to the breaking of the separate conservation law for the energy momentum tensors and enables a constant redistribution of heat between the left and right Weyl cones. This induced breaking of the separate conservation of the energy momentum tensors in a chiral electron system is a macroscopic form of the gravitational anomaly. We have planned such experiments and performed first test measurements, which make us optimistic that we can probe this consequence of the gravitational anomaly in the near future.

## 5. Magnetic Weyl semimetals

The first predicted Weyl semimetals were magnetic, namely the pyrochlore iridate $Y_2Ir_2O_7$ and $HgCr_2Se_4$ [39,40]. An overview with all refences about the research area is given in [41]. In Figure 6 a single crystals of important magnetic Weyl semimetal are displayed and the characteristic experimental methods are summarized. Angle resolved photoemission (ARPES) and Scanning Tunneling Microscopy (STM) allow for the direct monitoring of the bulk and surface electronic structure. For the ideal topological insulator, the surface state is a Dirac cone while the bulk electronic structure is gapped. In a magnetic TI the surface states are gapped, and properties such as axion insulator or quantum anomalous Hall can be observed. In magnetic Weyl and Dirac semimetals linear dispersion can be seen in bulk and Fermi arcs at the surfaces, if the material is magnetized. In other topological magnetic topological materials, the bulk spectra consist of nodal lines and more complex surface states are obtained as for example drum head states. Transport measurements with typical external stimuli such as electric and magnetic fields, light, temperature, pressure, strain are available for manipulating electronic properties of the magnetic topological materials. The classical anomalous Hall Effect (AHE) exists in nearly all ferromagnetic semimetals and metals. In magnetic topological materials the Berry curvature plays an important role. Magnetic Weyl semimetals are common: every crossing point in the band structure of a ferromagnetic centrosymmetric compound is related to nodal lines or Weyl points. An enhanced Berry curvature and a strong linear electromagnetic response leads to a large AHE and a large anomalous Nernst effect (ANE). The measurement set up of the Nernst effect is related to the AHE setup, instead of a current a thermal gradient is applied. The ANE is a transverse thermomagnetic effect, the anomalous transverse voltage is generated perpendicularly to the temperature gradient and the magnetization.

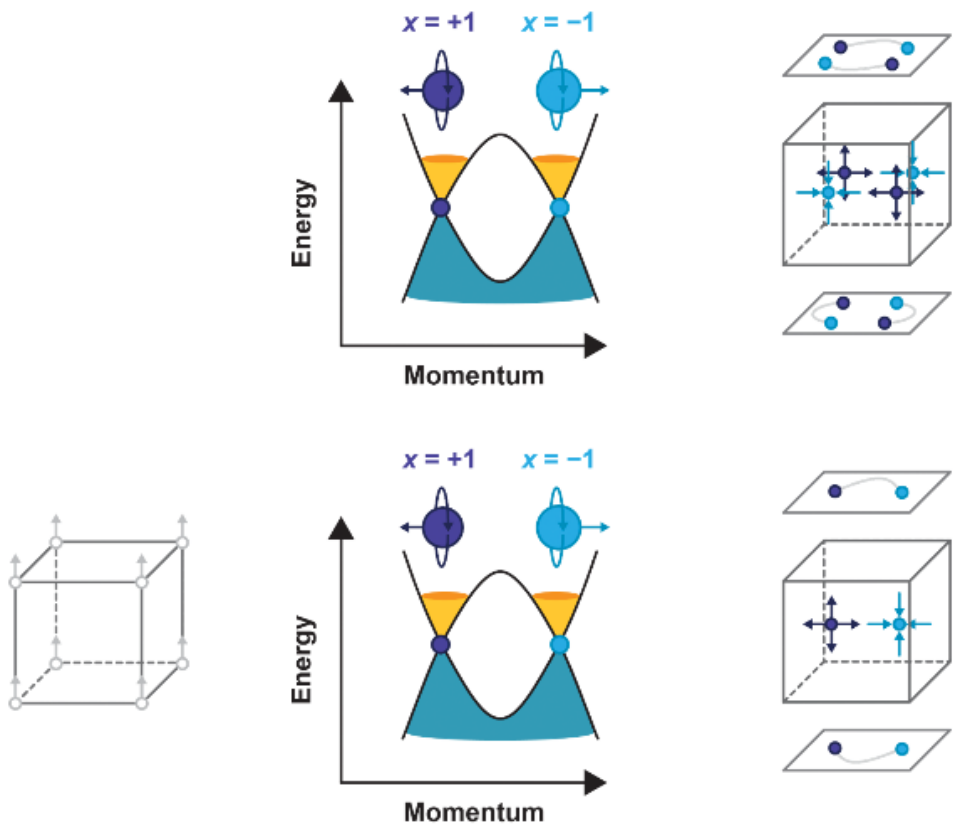

Fig. 6. Comparison of the simplified electronic structure of a non-magnetic Weyl semimetal and a ferromagnetic Weyl semimetal and the corresponding chiral Weyl points in the bulk and Fermi arcs on the surface of a crystal.

The chiral anomaly and the gravitational anomaly are also expected for magnetic Weyl semimetals. As an example, the chiral anomaly of the magnetic Weyl semimetal $Co_3Sn_2S_2$ is displayed in Figure 6 b [43]. However, similar to the chiral anomaly, circularly polarized light can induce an asymmetry between left- and right-handed chiral quasiparticles in nonmagnetic and magnetic Weyl semimetals. The light matter interaction with magnetic topological materials is an unexplored area, which we plan carefully investigate in near future.

In retrospect, this is a comprehensible development, since magnetic Weyl metals are close relatives of quantum Hall systems, which, as described above, represent the first topological solid-state systems. Quantum Hall systems always contain externally applied magnetic fields, i.e. in experiments, and therefore do not appear as a natural solid. One of the first ideas for a topological solid without an external magnetic field were two-dimensional crystals, which have a similar band structure as quantum Hall systems without a magnetic field, but are also ferromagnets. In such materials, the role of the external magnetic field is taken over by the intrinsic magnetization. Here, too, the topological invariant can be measured directly as a so-called quantum anomalous Hall effect — abnormal due to the lack of an external magnetic field. If such two-dimensional quantum anomalous Hall systems combine to form a three-dimensional solid crystal, a Weyl semimetal is created. Starting from a layered system of individual quantum anomalous Hall systems, as the coupling of the two-dimensional layers becomes stronger, the inverted band gap closes in the coupling direction and creates Weyl crossing points at the edge of the band structure, which move further and further into its center. In general, one can even say that every crossing point of electronic tapes in ferromagnetic materials is a Weyl point. In the simplest case, ferromagnetic crystals have at least four Weyl points Figure 7.

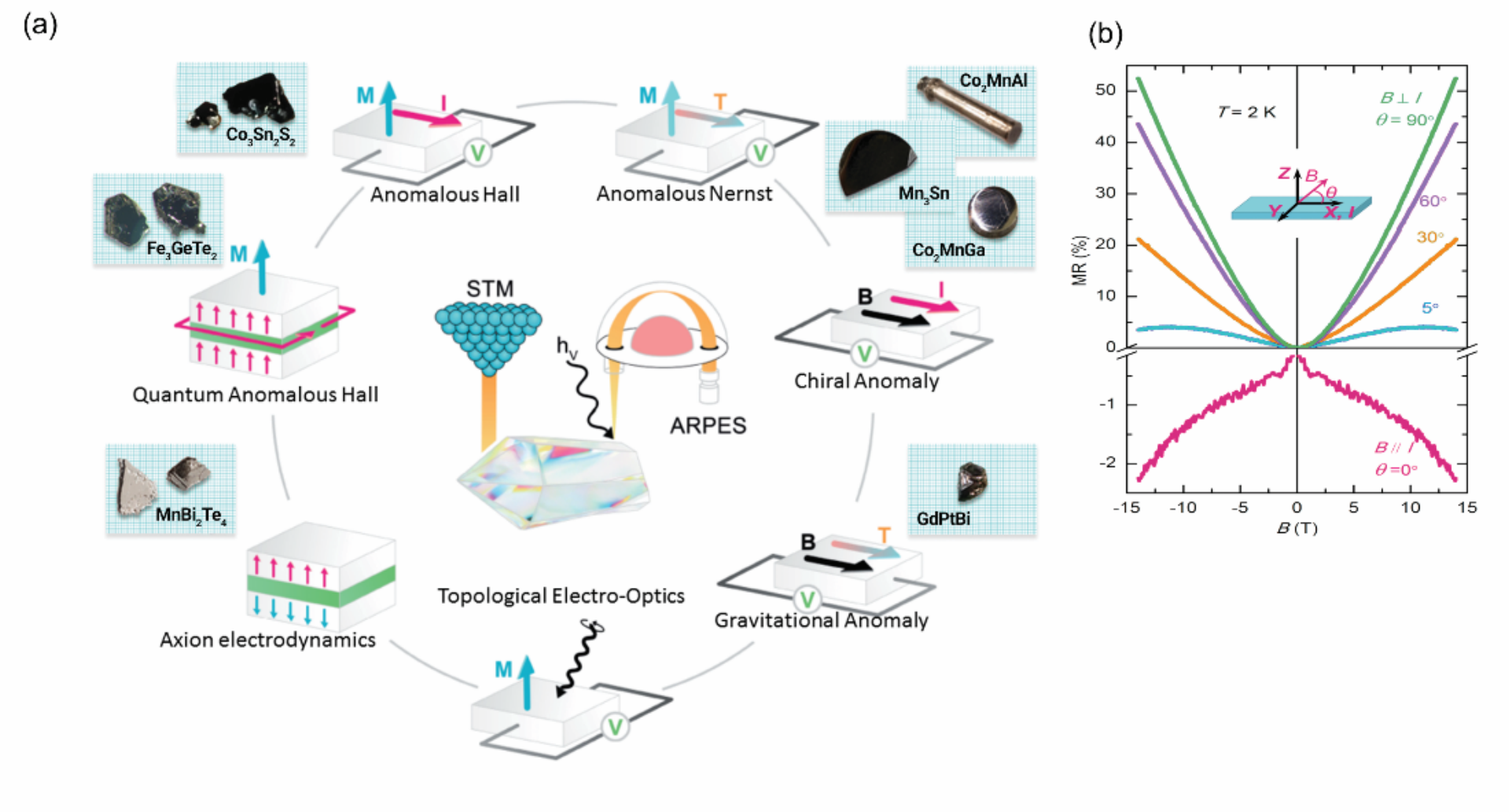

Fig. 7. Figure (a) according to [41]. In the outer circle, single crystals of the ferromagnetic Weyl semimetals $Co_2MnGa$, $Co_2MnAl$ and $Co_3Sn_2S_2$, the antiferromagnetic Weyl semimetals $Mn_3Sn$ and GdPtBi and the topological insulator $MnBi_2Te_4$ and the topological ferromagnetic metal $Fe_3GeTe_2$ are displayed. Physical investigation methods and characteristic properties of a magnetic Weyl semimetals are shown in the inner circle: Angle resolved photoemission (ARPES) and Scanning Tunneling Microscopy (STM) allow for the direct monitoring of the bulk and surface electronic structure. In a magnetic topological insulator some of the surface states are gap, and properties such as axion insulator or quantum anomalous Hall can be observed. In magnetic Weyl and Dirac semimetals linear dispersion can be seen in bulk and Fermi arcs at the surfaces, if the materials is magnetized. In other topological magnetic topological materials such as $Co_2MnAl$ and $Co_2MnGa$, the bulk spectra consist of nodal lines and more complex surface states are obtained as for example drum head states. Transport measurements with typical external stimuli such as electric and magnetic fields, light, temperature, pressure, strain are available for manipulating electronic properties of the magnetic topological materials. The classical anomalous Hall Effect (AHE) exists in nearly all ferromagnetic semimetals and metals. In magnetic topological materials the Berry curvature plays an important role. Magnetic Weyl semimetals are common: every crossing point in the band structure of a ferromagnetic centrosymmetric compound is related to nodal lines or Weyl points. An enhanced Berry curvature and a strong linear electromagnetic response leads to a large AHE and a large anomalous Nernst effect (ANE). The measurement set up of the Nernst effect is related to the AHE setup, instead of a current a thermal gradient is applied. The ANE is a transverse thermomagnetic effect, the anomalous transverse voltage is generated perpendicularly to the temperature gradient and the magnetization. The chiral anomaly is a smoking gun experiment for all Weyl semimetals. The gravitational anomaly is related to the chiral anomaly and should be observed in magnetic Weyl semimetals. A thermal gradient substitutes a current and a positive longitudinal magneto-thermoelectric conductance is measured for collinear temperature gradients and magnetic fields. The light matter interaction with magnetic topological materials is an unexplored area. Figure (b) according to [43]. The chiral anomaly for the ferromagnetic Weyl semimetal $Co_3Sn_2S_2$.

In antiferromagnetic compounds Weyl physics was recognized first [13,14], since an anomalous Hall effect (AHE) which is typical for ferromagnets was theoretically found based on an unusual Berry curvature. A common understanding was that the anomalous Hall effect is proportional to the magnetization. All antiferromagnets that have zero magnetization should not have an AHE. However, the non-collinear triangular antiferromagnetic arrangement in $Mn_3Z$ ($Z$ = Ge, Sn) [12] lead to a non-vanishing Berry curvature in momentum space and Weyl points at the Fermi energy. Shortly after the prediction a large AHE at room temperature was observed in $Mn_3Sn$ [13,44] and $Mn_3Ge$ [14]. However, it is known that numerous ferromagnetic Heusler compounds are still ferromagnetic even at room temperature and even up to over 900 °C. Therefore, it was not surprising that the first predicted ferromagnets were Co-Heusler compounds: $Co_2YZ$ ($X$ = V, Zr, Nb, Ti, Hf, $Z$ = Si, Ge, Sn) [45,46]. However, here the crossings are far above the Fermi energy and therefore not reachable in transport. With $Co_2MnAl$ [47,48] ferromagnets were predicted to be Weyl semimetals with Weyl points near the Fermi energy, with, as a consequence, giant anomalous Hall effects (AHE) due to the large Berry phase. In thin films of $Co_2MnAl$ a large AHE, in excellent agreement with theory, has already be found in [49]. $Co_2MnGa$ and $Co_2MnAl$ were found to be very large anomalous Hall values, in excellent agreement with the theory. The anomalous Hall angle of up to 12% in $Co_2MnGa$ is also promising [50]. The proof that these materials are magnetic Weyl semimetals was provided by electronic structural studies of $Co_2MnGa$. In this case, too, the expected surface conditions could be observed using ARPES [51]. Realizing the QAHE at room temperature would be revolutionary as it would overcome the limitations of many of today's data-based technologies that are affected by large electron scattering-induced power losses. This would pave the way for a new generation of quantum electronics and spintronics devices with low energy consumption.

In 2018 we realized an intrinsic hard magnetic Weyl semimetal with Weyl crossing points near the Fermi energy [43]. The Shandite crystals contain transition metals on a quasi-two-dimensional Kagome lattice. One of the most interesting candidates is $Co_3Sn_2S_2$, which has the highest magnetic order temperature within this family and in which the magnetic moments on the Co atoms are oriented in a direction perpendicular to the Kagome plane. Magnetic field-dependent electrical transport measurements indicate the chiral anomaly [43] and ARPES measurements clearly demonstrate the existence of magnetic

Weyl Fermions and very long Fermi arcs [52]. $Co_3Sn_2S_2$ shows a huge anomalous Hall effect up to temperatures of 150 K and a huge Hall angle overall, which indicates a Weyl semimetal that is still very close to layered anomalous quantum Hall systems. For a large Hall angle, two conditions must be met: first, a large Hall conductivity and, second, a small number of electrons. These conditions are met in Weyl semimetals in which the Weyl crossing points are close to the Fermi energy. In other words, the coupling of the Kagome planes in $Co_3Sn_2S_2$ is weak and it has a two-dimensional magnetic and electronic structure. In the compound, three cobalt atoms share a free electron. In fact, we have found that if you divide the measured anomalous Hall effect by the thickness of the crystal, you get a value that is roughly expected to be the quantum anomalous Hall effect per layer. We were also able to observe quantization of the edge states on cobalt edges of the material with scanning tunneling microscopy [53]. Subsequent band structure calculations actually showed the presence of Weyl nodes near the Fermi energy, calculations of the transport properties indicate a direct connection between the Weyl nodes and the increased anomalous Hall effect. In addition, strongly increased thermoelectric could be predicted and shown experimentally. In addition to the interesting quantum effects, magnetic Weyl semimetals also have a high potential in thermoelectric applications. In principle, the Fermi energy can be shifted in every ferromagnet (by substituting elements) so that the Weyl points are located at the Fermi energy. The observation of the quantum anomalous Hall effect at room temperature would enable novel computer technologies including quantum computers. To realize this possibility, our strategy is to look for quasi-two-dimensional magnetic materials with topological band structures — that is, three-dimensional crystals that look almost like individually layered anomalous quantum Hall systems — and to apply these materials as monolayers or very thin films synthesize. So far, however, no magnetic materials are known that could lead to a quantum anomalous Hall effect with a higher temperature. With a magnetic transition temperature of 150 K in $Co_3Sn_2S_2$ we are still a long way from potential room temperature effects.

## 6. Chiral electrons beyond Weyl semimetals

As already mentioned, electrons in conventional Weyl semimetals are described by relativistic equations and were originally proposed in the field of high energy physics. But there are also solid-state crystals, in which chiral electrons have no corresponding analogues in high-energy physics and are therefore described as 'New Fermions' [15-17,54]. In contrast to the vacuum considered in high energy physics, electronic (quasi)particles in condensed-matter systems are not constrained by Poincare symmetry. Instead, they must only respect the crystal symmetry of one of the 230 space groups. Hence, there is the potential to find and classify free fermionic excitations in solid-state systems that have no high-energy counterparts. While Dirac Fermions are four times degenerate and Weyl Fermions are degenerate twice, the new Fermions even show six- and eight-times degeneracy. Particular examples, in which inversion symmetry is broken, are crystals with a chiral crystal lattice. Such chiral crystals are examples, in which the concept of chirality appears in real space. In this context, chirality refers to the property of these crystals that their atoms follow a spiral, step-like pattern as in the biological systems, such as DNA. While the staircase rotates clockwise in one system, it rotates counterclockwise in the opposite system (Figure 8), but both systems have identical composition. These systems, called “enantiomers”, are mirror images of each other (Figure 8 a). Chiral crystals with heavy elements, topological chiral Fermions are of particular interest. In the family of the B20 structure type, the $P2_13$ (198) space group (Figure 8 a, shows both enantiomers), a new type of electron, the so-called Rarita-Schwinger Fermions, was confirmed. The two band crossing points in crystals with broken inversion symmetry, are at different energies (Figure 8 b and d) and show six- and four-times degeneracy. This leads to several notable properties, including: a huge quantized circular photo-galvanic current [56-58], a chiral magnetic effect, and other novel transport and optical effects not observable in Weyl semimetals. Various candidate materials such as PdGa were selected for ARPES studies because the spin orbit coupling is very large and both enantiomers could be synthesized from this compound. The complex band topology of the two enantiomers leads to Fermi arcs (Figure 8 c and e), [16,54,55,63-65] which, however, are also mirror images of each other (Figure 8 e). The high spin-orbit coupling is responsible for a strong splitting of the bands so that the experimental resolution was sufficient to confirm the four bands representative of the topological number of the Rarita-Schwinger Fermions. The experiments also show the expected band degeneracies at the highly symmetrical points of the Brillouin zone with the crossing points at different energies. Fermi arcs that run across the entire Brillouin zone and are also chiral, with different chirality in the lattice structure for the two enantiomers, are shown in Figure 8 e.

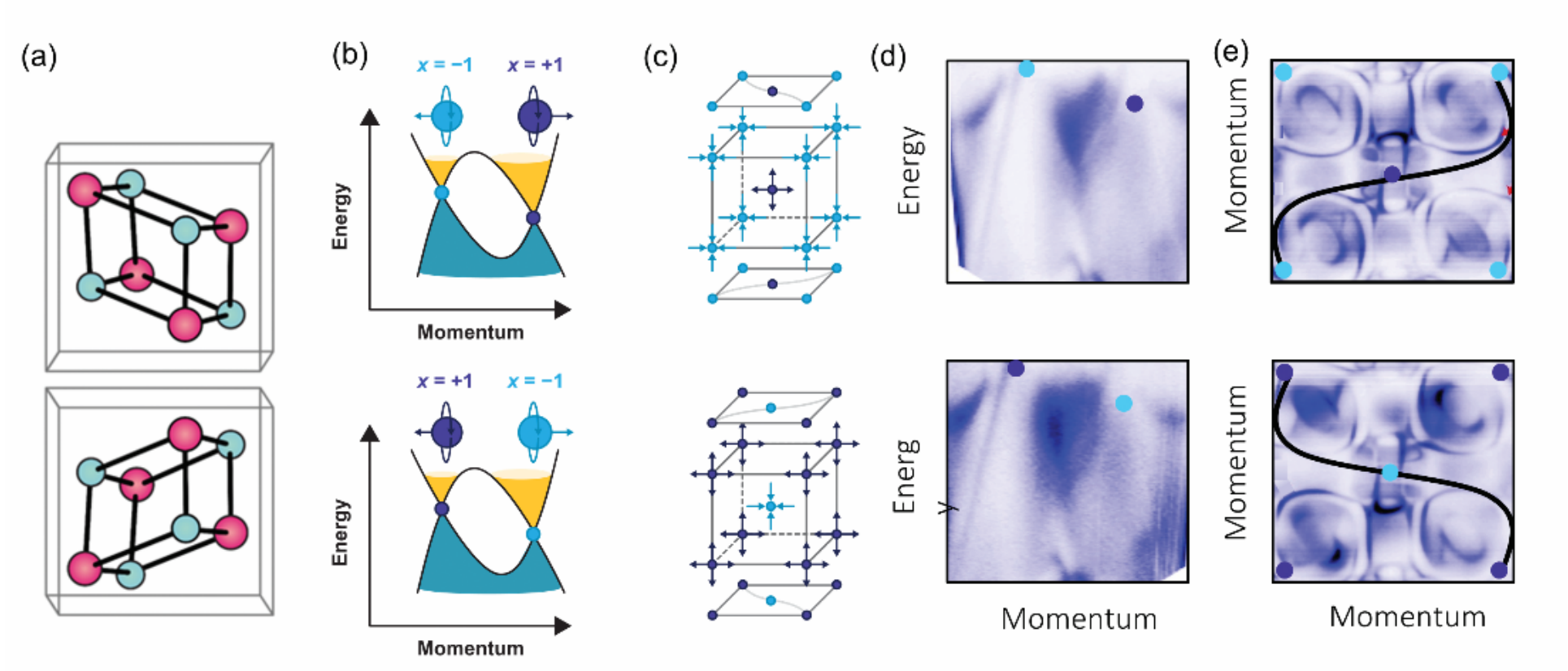

Fig. 8. Figure according to [16]. Experiments on chiral Weyl semimetals. (a) Sketch of the atomic structure of PdGa (red spheres represent Ga atoms and silver spheres represent Pd atoms), enantiomer A and enantiomer B. (b) Sketch of the electronic structure of enantiomer A and B, and (c) its distribution of the Weyl crossing points and the corresponding chiral surface states. (d) Images of the angle-resolved emission spectroscopy of the interior of PdGa enantiomer A and enantiomer B. (e) Images of the angle-resolved emission spectroscopy of the surface of PdGa enantiomer A and enantiomer B.

In general, chiral inorganic or organic materials show exceptional optical properties and crystallize in one of the 65 Sohncke symmetry space group [59-61]. If the space group contains only proper operations, the crystal is chiral and obey a well-defined handedness.

There are more than 100 000 non-biological chiral crystals. Of all inorganic compounds in the inorganic database ICSD approximately 20% are chiral [59]. Beside the structural chirality in materials itself, materials can host chiral quasiparticles and excitations such as chiral electrons, chiral photons, chiral magnons, chiral plasmons, chiral spins etc. . In a few cases it is speculated whether there is a relationship between the handedness of structure and handedness of the chiral objects such as Skyrmions, even on different length scales (atomic distances of Å and Skyrmions nm–μm) [62]. The underlying correlation between a chiral lattice structure and the chiral quantum number of the electrons remains largely unexplored to date. However, it has been suggested that while the band structure in the bulk of the two enantiomers is identical, the chirality of the electrons and the Berry curvature behave like image and mirror image.

## 7. Weyl semimetals with strongly interacting electrons

While Weyl systems with free electrons are experimentally well investigated today, materials with strongly interacting electrons are largely unexplored. By switching on strong interactions between electrons with vibrations of the crystal lattice, so-called phonons, a charge density wave can be induced in Weyl semimetals, which connects the two Weyl crossing points and opens an energy gap at them. The resulting material is a so-called axion isolator (Figure 9 a). A charge density wave is the energetically preferred ground state of a strongly coupled electron-phonon system in certain quasi-one-dimensional metals and semi-metals at low temperatures. It is characterized by a gap in the dispersion of the free electrons and by a collective metallic mode which, like a superconductor, is formed by electron-hole pairs. The electron distribution and the position of the lattice atoms are periodically modulated in real space with a period that is greater than the original lattice constant (Figure 9 b). The electric current-carrying collective mode, the so-called phason, is a solid-state version of the axion particle, which is traded as a possible candidate for dark matter in high-energy physics, but has not yet been observed. In most real solids, the phase is bound to impurities. Therefore, it can only "slide" freely over the grid and contribute to the flow of electrical current when a certain electrical threshold field is applied (above which the electrical force overcomes the binding forces). The resulting conduction behavior of the solid is strongly non-linear — as in the case of a diode, for example — and the electric current increases as the electric field increases. The signature of an "axionic" charge density wave is then, corresponding to the chiral anomaly in a Weyl semimetal with free electrons, a positive longitudinal magnetic field-dependent electrical conductivity. However, like its elementary particle analogue, this axionic charge density wave was not experimentally demonstrated until recently.

Only in 2019, we succeeded in measuring a large positive contribution to the magnetic field-dependent electrical conductivity of the phason [66,67] of the charge carrier density wave in the Weyl semimetal $(TaSe_4)_2I$ for collinear electric and magnetic fields (Figure 9 d).

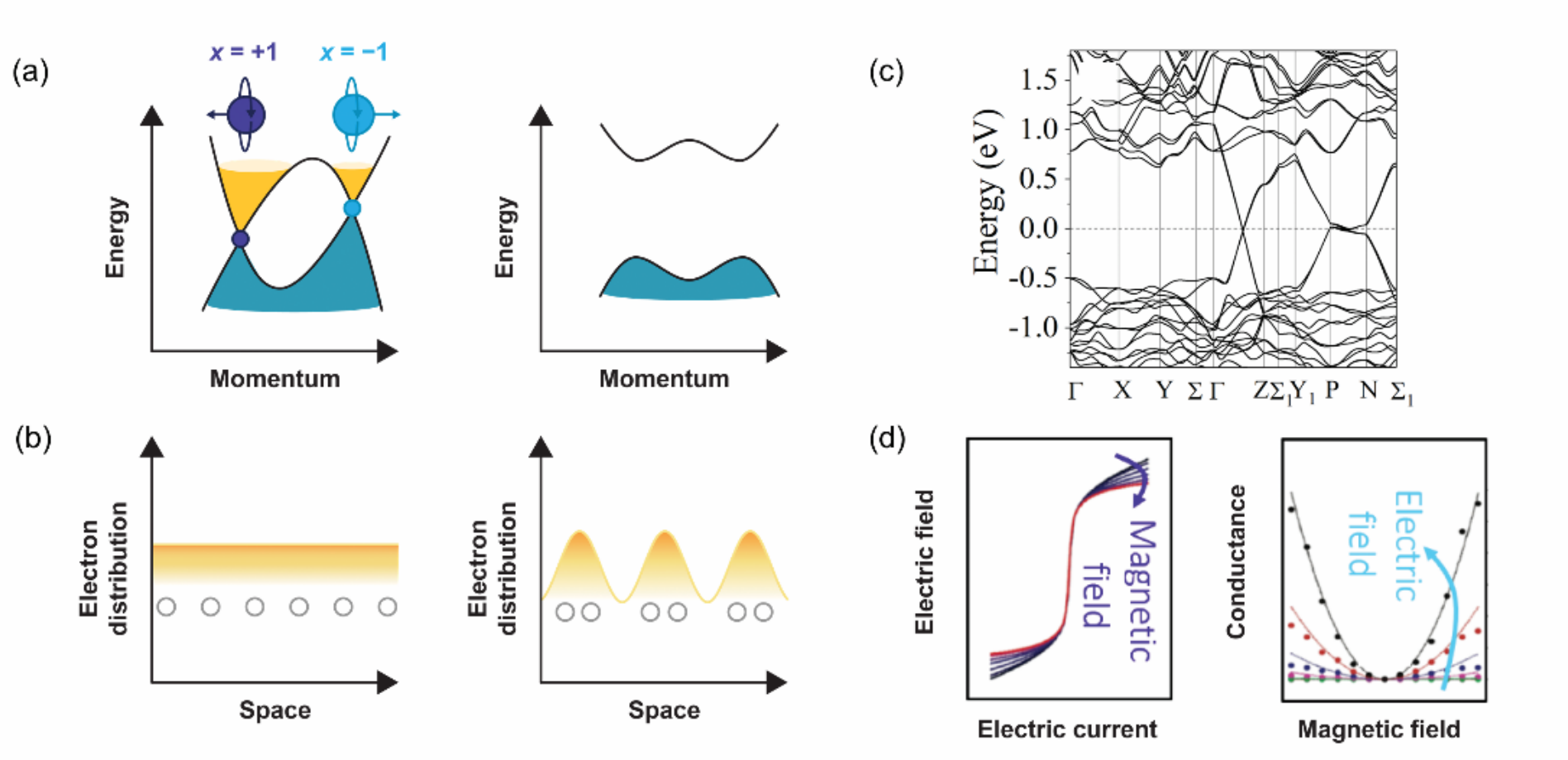

Fig. 9. Axion charge carrier density wave in a Weyl semimetal. (a) Band structure of the free electrons (top) and electron distribution in real space for a Weyl semimetal. (b) Periodic modulation of the electron distribution (charge carrier density wave) through strong electron-phonon interaction in real space creates a band gap at the Weyl crossing points. The crystal lattice is also modulated. The collective mode of the charge carrier density wave, the phason, is an axion. Figure (c) according to [68], the band structure of the Weyl semimetal $(TaSe_4)_2I$. (d) Transport experiment on the axion charge carrier density wave. In the case of high electric fields, where that of the phason dominates the electric current flow, the current is increased by a magnetic field applied parallel to the electric field. The magnetic field dependence of the longitudinal conductivity agrees with the anomalous transport of an "axionic" charge density wave.

We were able to show that this positive contribution results from the anomalous axionic contribution of the chiral anomaly to the electrical current flow and that this is linked to the parallel alignment of the electrical and magnetic fields (Figure 9 d). By rotating the magnetic field, we showed that the angular dependence of the phason conductivity is consistent with the anomalous transport of an axionic charge density wave. Our results show that it is possible to find experimental evidence for axions in highly correlated topological systems of condensed matter that were previously not found in any other context.

## 8. Outlook

It is intriguing that nature provides us with Chirality and symmetry breaking at several levels, ranging from the most fundamental standard model in particle physics to homochirality in biomolecules. While many hypotheses have been provided, and intriguing experiments have been performed, most fundamental questions on chiral symmetry breaking remain largely unanswered. Already Pasteur conjectured that spatial crystal chirality (which he called 'dissymmetry') in living systems [68-70] like as in DNA molecular may be induced by some universal chiral force or influence in nature. The chiral anomaly, in principle provides such an "influence" on a fundamental quantum level, but whether there is such a connection between the chiral quantum number and spatial structures is not known to date in the high energy context. Key insights may arise from other areas of physics, e.g. an analogy has been found in solid-state systems hosting Weyl

Fermions [22,30-32] but also can exhibit chiral crystal structures. In fact, we believe, the topological materials discussed above might allow to test such a connection in future experiments.

On the one hand, the chiral and gravitational anomalies have been observed on the electron quasi-particle level in Weyl semimetals and on the other hand, we have observed first signatures of an imprint of the chiral crystal structure in non-centrosymmetric topological semimetals on the distribution of the chiral quantum number of their electron states in momentum space. Specifically, one possible root to explore this connection further in Dirac, Weyl and new Fermion semimetals might be to externally twist or apply torsional torque to materials. In particular, the chiral anomaly in the quasiparticle framework could be studied in materials with structural chirality and even in twisted 2D materials [71-73]. Controlling and promoting chirality in real and momentum space could be a future direction in topological material science. Light matter interaction is another important and unexplored direction which may couple topological band structures and spatial chirality. And as Frank Wilczek, who named the axion [74], said very recently: “If we know that there are some materials that host axions, well, maybe the material we call space also houses axions [75].”

On the other hand, molecular catalysis and the selective adsorption of helical molecules can happen on surface states, where high mobilities, spin momentum locking and chirality might be important criteria for efficiency and selectivity of molecular handedness. In this regard, topological surface states may favor heterogeneous catalysis processes such as the hydrogen evolution reaction (HER). Recently, we started to perform experiments that can potentially build a bridge between chiral surface states and chiral molecules. We observed for example that the efficiency for HER on PdGa, PtAl and PtGa metals (space group of $P2_13$ (#198)) on surfaces with a long topological Fermi arc (Figure 10) is unexpectedly high, with turnover frequencies as high as 5.6 and 17.1 $s^{-1}$ and an overpotential as low as 14 and 13.3 mV at a current density of 10 mA $cm^{-2}$. These chiral crystals outperform commercial Pt and nanostructured catalysts [76,77]. First experiments show evidence that these materials allow enantiospecific control and the asymmetric on-surface synthesis of homochiral molecules [78].

In parallel, the so called CISS effect, that is chirality-induced spin selectivity (CISS) [79-81] has recently attracted attention, but is still not fully understood. We believe at the core of all these open problems is the question “how the information of a chiral quantum number transfers to a structural chirality”. Possible options under discussion are the interplay of spin-orbit coupling, spin momentum locking and a geometric phase similar to the Berry phase in real space and brings chiral molecules close to topology. As explained above, we have made the first steps to understand this connection. Although our results are in part preliminary and cannot yet be generalized, they give already such strong indications for the connection of chirality in momentum and real space that we are excited about working on topology and chirality in the future!

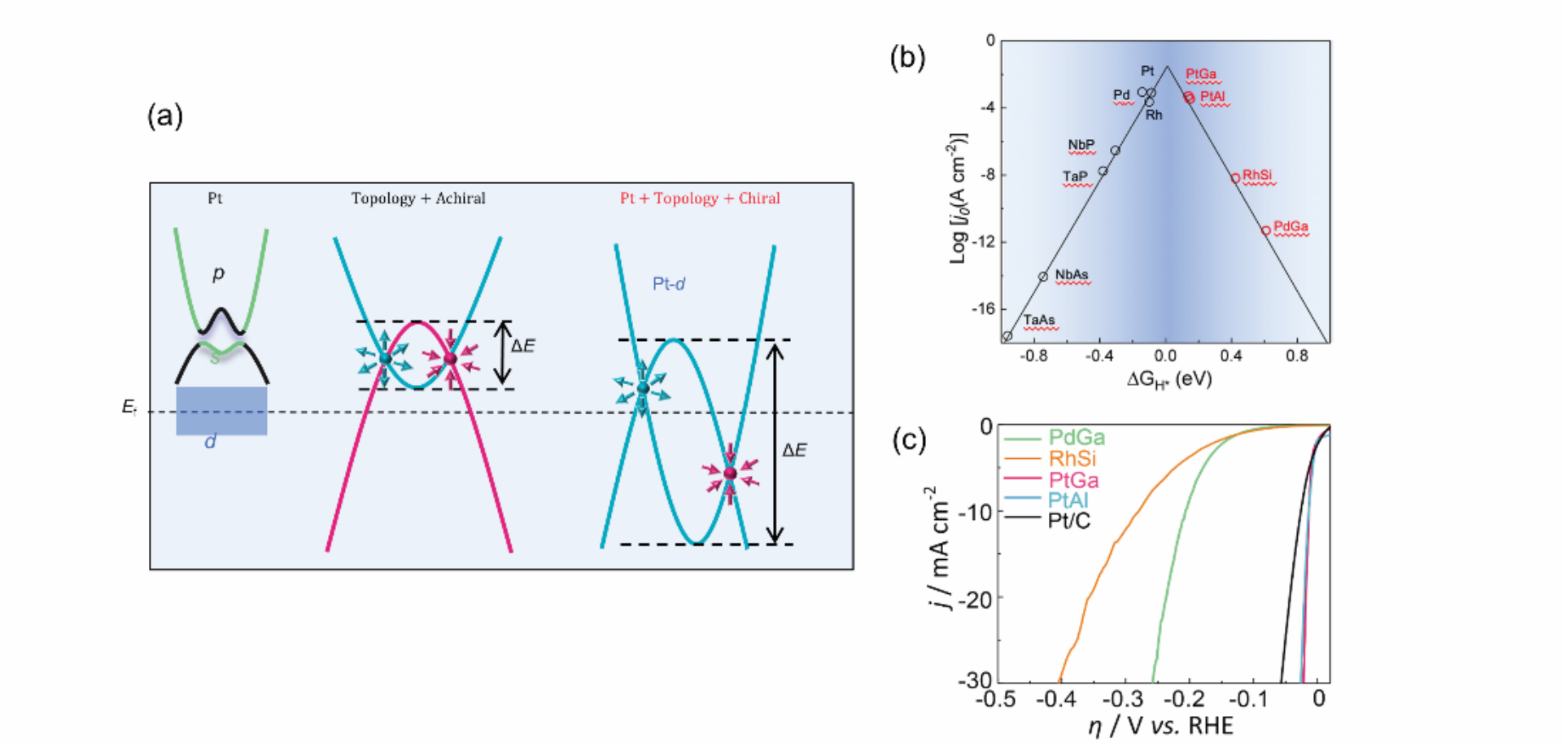

Fig. 10. Figure according to [76]. Hydrogen evolution reaction with Pt-group metal based chiral crystals. (a) Illustration of the band inversion mechanism and topologically nontrivial energy window in pure Pt, non-chiral topological semimetals, and chiral B20 compounds. (b) Volcano plot of PtAl, PtGa, RhSi, PdGa, the related metal catalysts and Weyl semimetals NbP, TaP, NbAs, and TaAs are also presented for comparison. (c) Polarization curves of the chiral crystal catalysts: PdGa, RhSi, PtAl, PtGa, and 20% Pt/C catalysts.